**PAPER**

# What Makes Zeolitic Imidazolate Frameworks Hydrophobic or Hydrophilic? Impact of Geometry and Functionalization on Water Adsorption[†]

Aurélie U. Ortiz,[a] Alexy P. Freitas,[a] Anne Boutin,[b] Alain H. Fuchs,[a] François-Xavier Coudert[a,*]

We demonstrate, by means of Grand Canonical Monte Carlo simulation on different members of the ZIF family, how topology, geometry, and linker functionalization drastically affect the water adsorption properties of these materials, tweaking the ZIF materials from hydrophobic to hydrophilic. We show that adequate functionalization of the linkers allows one to tune the host–guest interactions, even featuring dual amphiphilic materials whose pore space features both hydrophobic and hydrophilic regions. Starting from a hydrophobic initial material (ZIF-8), various degrees of hydrophilicity could be obtained, with a gradual evolution from a type V adsorption isotherm in the liquid phase, to a type I isotherm in the gas phase. This behavior is similar to what was described earlier in families of hydrophobic all-silica zeolites, with hydrophilic "defects" of various strength, such as silanol nest or the present of a extra-framework cations.

## I. Introduction

Much attention has recently been focused on metal–organic frameworks (MOFs), a wide class of microporous materials that garner a lot of interest for their potential applications in the fields of separation, catalysis, strategic gas capture and storage, and drug delivery.[1,2,3,4,5,6] Zeolitic imidazolate frameworks (ZIF) are a subclass of metal–organic frameworks (MOF) that feature imidazolate linkers bridging metal centers to form three-dimensional porous crystalline solids isomorphous to zeolitic frameworks.[7,8,9,10] ZIFs have recently gained considerable attention for their potential applications because they inherit desirable qualities from both worlds: the tunable porosity, structural flexibility and the functionalization of the internal surface of the MOFs, as well as the thermal and chemical stability of the zeolites. Moreover, the similarities between the metal–imidazolate four-fold coordination chemistry and the corner-sharing $SiO_4$ tetrahedra from which zeolites are built mean that many ZIF topologies can potentially be synthesized. Indeed, over 100 different ZIF structures have been reported so far, and theoretical[11,12] as well as experimental[13] investigations of the relative stabilities of other polymorphs yet to be synthesized indicate that many of them have a relatively low enthalpy of formation, and should be accessible under mild synthesis conditions. The ZIF structures reported in the literature so far demonstrate large gamut of attractive structural and physicochemical properties, with great potential for applications in $CO_2$ capture,[14] sensing,[15] encapsulation and controlled delivery,[16] and fluid separation.[17,18,19,20]
While much attention has been paid to the adsorption of gases and liquids in ZIF materials, including carbon dioxide, hydrogen, methane, longer alkanes, and alcohols,[21] there has been relatively little information reported on water adsorption in this large family of materials. Küsgens et al.[22] initially reported the water vapor adsorption isotherm at 298 K for ZIF-8, a Zn(2-methylimidazolate)$_2$ porous framework with sodalite (SOD) zeolitic topology. This isotherm demonstrated the strong hydrophobic character of this material and its remarkable hydrothermal stability; these findings were later corroborated by Cousin Saint Remi et al.[23] Guillaume Ortiz et al.[24] later investigated the energetic performance of ZIF-8 in high-pressure liquid water intrusion–extrusion experiments. They showed reversible intrusion–extrusion cycles for liquid water in the 15–35 MPa range, with a shock-absorber behavior and energetic yield of 85%, close to pure silica zeolites. Zhang et al.[25] studied the impact of ZIF-8's hydrophobicity and low water uptake on its ethanol/water separation capabilities, concluding that ZIF-8 presented a good adsorption selectivity.
But ZIF-8 was not the only material of the ZIF family for which water adsorption was studied. Lively et al.[26] presented the isotherm of water adsorption, again in the gas phase, on ZIF-71 (whose linker is 4,5-dicholoroimidazolate): the material is again hydrophobic, with very little water uptake at $P < P^0$. Moreover, Zhang et al. recently published an experimental study of water and C1–C4 alcohols adsorption in ZIF-8, ZIF-71 and ZIF-90 (whose linker is imidazole-2-carbaldehyde).[27] They showed that ZIF-90 presented a stepped isotherm in the gas phase, with water uptake around $P/P^0 \sim 0.4$. Biswal et al.[28] reported the adsorption and desorption isotherms of water in CoNIm, a cobalt-based ZIF with 2-nitroimidazole linkers with RHO topology. These adsorption isotherm is characteristic of a hydrophilic material, with an additional step near $P/P^0 \sim 0.6$. However, while the CoNIm appears stable in presence of water (as established by powder X-ray diffraction patterns), the desorption shows nonreversible behavior and the adsorbed water could not be fully evacuated.

On the theoretical side, a few groups have used molecular simulations in order to assess the water adsorption in various ZIF frameworks. Nalaparaju et al. published in 2010 a study of water adsorption in hydrophobic ZIF-71, as well as a hydrophilic Na-rho-ZMOF (a MOF with anionic framework and $Na^+$ extraframework cations).[29] Their results confirmed the hydrophobic nature of ZIF-71, with no water uptake in the gas phase and a type V isotherm with large hysteresis (8–25 kPa) for liquid water. Finally, Amrouche et al.[30] recently studied the low-pressure adsorption of water in a series of ZIFs, using the ideal heat of adsorption of water and Henry's constant[31] as descriptors of hydrophobicity or hydrophilicity.

In this paper, we used molecular simulation to investigate the water adsorption properties of 7 ZIFs, in order to understand and rationalize the influence of topology, geometry, and linker functionalization on the hydrophobic or hydrophilic nature of the porous solid and the water–ZIF interaction strength. In particular, we show how very small changes in geometry at fixed topology and chemical nature, or very small changes in chemistry at fixed topology and geometry, can have a drastic impact on water adsorption properties. We then compare these results to the existing body of knowledge obtained on water adsorption in hydrophobic and hydrophilic zeolites, drawing a parallel between these two related families of nanoporous materials.

## II. Systems and simulation methods

### 1. ZIFs studied in this work

In this work, we studied water adsorption into seven materials of the ZIF family; some of them are experimentally known structures, some of them are hypothetical structures. The first structure is the widely-studied ZIF-8, a polymorph of $Zn(mim)_2$ with SOD zeolite topology (*mim* = 2-methylimidazolate).[7] ZIF-8 presents large spherical cages of diameter 11.6 Å, called sodalite cages, separated by 6-ring windows of small aperture (~3.4 Å diameter as determined from the crystallographic structure). Its structure is presented in Figure 1. It has a good separation performance for strategic gas mixtures such as $CO_2/CH_4$, and high thermal, mechanical and chemical stability. It is available commercially, and can be readily assembled into membranes or form thin films.

We also performed simulations on six other experimentally known ZIFs with SOD topology but different functional groups: ZIF-90,[32] where the linker (*ica* = imidazolate-2-carboxyaldehyde) has an carboxyaldehyde group instead of a methyl group; SALEM-2,[33] in which the linker is an unfunctionalized imidazolate;[34] ZIF-Cl,[35] with a chloro-functionalized linker (*cim* = 2-chloroimidazolate); ZIF-7,[36] a $Zn(bim)_2$ polymorph featuring a distorted SOD topology (*bim* = benzimidazolate). We also included two hypothetical structures of $Zn(nim)_2$ (*nim* = 2-nitroimidazolate): these have been observed experimentally as cobalt-based $Co(nim)_2$ phases, under the name of ZIF-65,[14] with both SOD and RHO topology.[28] In order to maintain some consistency in the family of materials studied in this work, and to focus our comparison on the effect of linker functionalization and topology on water adsorption (rather than metal center), we thus considered the two hypothetical equivalent zinc-based materials, $Zn(nim)_2$ (SOD) and $Zn(nim)_2$ (RHO), which for simplicity we will henceforth call ZIF-65 (SOD) and ZIF-65 (RHO).

All ZIFs studied are of SOD topology, except for ZIF-65 (RHO), which is of RHO topology. The RHO zeolite topology is a very open net, with highly accessible internal pores. It is built from body-centered cubic arrangement of truncated cubo-octahedra (colored in orange in Figure 1), or α-cages, linked via double 8-rings (in green in Figure 1). All the frameworks studied and their unit cell parameters are reported in Table 1.

For SALEM-2 and ZIF-Cl, whose experimental crystalline structures are not available in the literature, we manually constructed the starting structure from that of the isostructural ZIF-8, and then performed energy minimization on both unit cell parameters and atomic positions by quantum chemistry calculations at the density functional theory (DFT) level with localized basis sets, as implemented in the CRYSTAL09 code.[37] The protocol followed was identical to that validated in recent work on various metal–organic frameworks:[38] the B3LYP exchange–correlation function[39] with empirical correction for dispersive interactions as proposed by Grimme in 2006 (the so-called "D2" correction).[40] All electron basis sets were used for all the atoms involved: 6-311G(d,p) for H, C, and O,[41] 86-411d31G for Zn.[42] The resulting structures are also included in the Supplementary Information.

### 2. Interaction potentials

The water molecules were modeled by the rigid, non-polarizable TIP4P model,[43] featuring three electrostatic charges and a single Lennard-Jones center. All the ZIF structures were considered rigid, which is only valid as a first approximation: it is known, for example, that the ZIF-8 structure features some local flexibility of its structure by a "linker swing" motion. However, the importance of flexibility on adsorption properties has so far mostly been observed at cryogenic temperatures, which justifies the approximation employed here. Moreover, flexibility of ZIFs other than ZIF-8 is still a completely unexplored area.

ZIF–water interactions were described by a classical force field including the repulsion-dispersion energy, modeled by Lennard-Jones 6-12 potentials, and Coulombic interactions, modeled by point charges on all atoms of the ZIF structures. Both the Lennard-Jones parameters of the ZIFs and the atomic point charges were taken from the earlier work of Amrouche et al.,[44] who optimized Lennard-Jones parameters based on the transferrable Universal Force Field (UFF) and Electrostatic Surface Potential (ESP)-fitted atomic charges from DFT calculations on each ZIF structure. This forcefield for ZIFs and ZIF–guest interactions has already been extensively tested.[44,45,46] All parameters of the forcefields used are detailed in the Supplementary Information.

### 3. Grand Canonical Monte Carlo simulations

Water adsorption in ZIFs have been simulated by means of forcefield-based Monte Carlo simulations in the grand canonical ensemble. For each material, series of GCMC simulations were performed at various values of water chemical potential, in the gas as well as in the liquid phase, first increasing in value (adsorption branch), then decreasing from the highest point (desorption branch). The chemical potential of water was then related to water pressure by the same $\mu(P)$ relation as in ref. 47, with the saturation pressure of the TIP4P model being $P^0 = 3.8$ kPa at 300 K. All Monte Carlo runs were performed at 300 K. Periodic boundary conditions were used and long-range electrostatic interactions were taken into account using the Ewald summation technique. In order to improve the efficiency of the calculations, electrostatic and repulsion–dispersion interaction energies between the rigid MOF framework and adsorbed water molecules were precomputed on a grid for each material (with a grid mesh of 0.1 Å) and stored for use during the simulation. Each simulation consisted of 100 million Monte Carlo steps, of which 50% were insertion/deletion moves (with preinsertion and the orientational bias), 25% were molecule translations and 25% were molecule rotations.

## III. Results and discussion

### 1. ZIF-8 and other hydrophobic ZIFs

First, we present the adsorption-desorption isotherms of water in ZIF-8 at 300 K (Figure 2). We observe that the ZIF-8 material does not adsorb any water in the gas phase, a clear sign of the established hydrophobic nature of the solid.[27] At higher pressure, way above the water model saturation pressure (3.8 kPa at 300 K), in the liquid phase, the isotherms exhibits a step and the water saturation uptake is around 80 molecules per unit cell (29.1 mmol/g), in very good agreement with water intrusion experimental results from ref. 24 (27.8 mmol/g). Moreover, the type V isotherms[48] presents a wide hysteresis loop ranging from 15 MPa to 140 MPa. There is quite good agreement on both the pressure range and in particular the position of the desorption branch of the isotherm (usually considered closer to the thermodynamic equilibrium). The later was reported in ref. 24 to happen at 20 MPa, close to our value of 15 MPa.

However, the width of the hysteresis loop cannot be directly compared with experimental results for water intrusion from Guillaume Ortiz et al.,[24] because they are controlled from entirely different physical phenomena. The existence of the hysteresis in both cases is due to the metastability of the empty and filled state of the porous material in a certain pressure range.[49] However, the extent of the hysteresis within this range of metastability is dictated by different factors for experiments and GCMC simulations. In the first case, the issue is one of kinetics and is highly dependent on the measurement setup (scan rate, pressure increments, etc.). For simulations, the extent of the hysteresis is directly determined by the convergence of the Monte Carlo algorithm, which in turns depends on number of steps performed but also type of MC moves and biases used.

The hydrophobic nature of ZIF-8 is also clearly visible on the heat of adsorption (red curve in Fig. 2), and on the adsorption enthalpy at zero loading (Table 2). The heat of adsorption for the first adsorbed molecules is ~20 kJ/mol, which is much lower that the bulk vaporization enthalpy of water (44 kJ/mol for our TIP4P water model at 300 K). After water intrusion (at $P > 140$ MPa), the heat of adsorption is around 52 kJ/mol, somewhat larger than bulk vaporization enthalpy, indicating a dense adsorbed phase with strong water–water interactions. We further investigated the structure of the water adsorbed inside the pores of the ZIF-8 material. In Figure 3, we reported the O–O and O–H radial distribution functions (RDF) for water adsorbed in ZIF-8 at saturation ($P = 250$ MPa). We find a strong ordering of water molecules at short distance, with a marked O–O first peak at 2.8 Å, a distance identical to that of bulk water. Unlike liquid water, there is no longer-range order (no second peak) because of excluded volume effects due to the material.[50] The O–H RDF shows two clear peaks characteristic of hydrogen bonding, with O–H distances of 1.9 Å and 3.2 Å (second hydrogen atom of a H-bonded neighbor). The position of both peaks is again very close to those of bulk water. We thus conclude that the structure of the water adsorbed in the ZIF-8 pores is similar to bulk liquid water.

Next, we were interested in the effect of small changes in linker functionalization on the water adsorption properties. Figure 4 presents the adsorption-desorption isotherms of liquid water in SALEM-2 and ZIF-Cl, two ZIFs isostructural to ZIF-8 whose linkers are the imidazolate and 2-chloroimidazolate anions, respectively. In both cases, the small and local change in linker functionalization induces a small effect on the adsorption–desorption isotherms and heats of adsorption. The materials retain their hydrophobic character, like ZIF-8, with type V isotherms with wide hysteresis loops. We can note that the hysteresis in the case of ZIF-Cl is slightly smaller than ZIF-8 and SALEM-2. Finally, the amount of water adsorbed at saturation is similar for ZIF-8 and ZIF-Cl (~78 molec./u.c., or 24.2 mmol.g$^{-1}$), owing to the comparable size of the methyl and chloride groups. In contrast to these two materials where water can only enter one type of aperture (6-ring windows), SALEM-2 presents additional accessible porous volume in the center of the 4-ring windows because of the smaller H atoms (compared to the –CH$_3$ and –Cl groups of ZIF-8 and ZIF-Cl). This allows the adsorption of more water molecules (~90 molec./u.c., or 37.6 mmol.g$^{-1}$). This observation is consistent with the results of Karagiaridi et al.[33] showing that SALEM-2 can accommodate larger guest molecules than ZIF-8.

### 2. Hydrophilic ZIFs: ZIF-65 (RHO) and ZIF-90

In a second stage, we turned our attention to hydrophilic ZIFs, starting with the only such ZIF from which experimental adsorption is available. The CoNIm (RHO) is a Co-imidazolate based ZIF with RHO topology synthetized by Biswal et al.[28] who reported the high water stability of the RHO topology material compared to that of the CoNIm (SOD) material with sodalite topology (also known as ZIF-65[14]). We studied the material in its hypothetical Zn-based variant, which we call ZIF-65 (RHO) by analogy to ref. 14 (see section II.1). Figure 5 presents the adsorption and desorption isotherms of water in ZIF-65 (RHO), in the gas phase. While the metal ion is not the same in both materials (Zn$^{2+}$ vs. Co$^{2+}$), we can still perform a qualitative comparison with the experimental isotherms (ref. 28), because the exact nature of the (unexposed) divalent metal ion is not expected to play a major role in adsorption properties.

In contrast to the isotherms of the ZIF-8 and other hydrophobic ZIFs, the simulated adsorption isotherms of ZIF-65 (RHO) presents a significant uptake of water at low pressure ($P \leq 2$ kPa), and reaches saturation with a vertical step at pressure near $P°$. This is in qualitative agreement with the experimental results from ref. 28; however, the experimental desorption branch shows nonreversible behavior indicating that some water molecules could not be fully desorbed.

Looking at the simulation isotherm in detail, the first part of the adsorption isotherm ($P \leq 2$ kPa), which is of type I, corresponds to the adsorption of water molecules inside the double 8-ring connecting the α-cages ($P \leq 2$ kPa). The vertical adsorption steps at 4 kPa is linked to water adsorption inside the spherical cavities (α-cages) of the material. The adsorption-desorption isotherms thus present a vertical step with wide hysteresis, characteristic of a hydrophobic pore surface. The saturation uptake amount corresponding to the water intrusion is around 450 molecules per unit cell due to the high porosity of the material with RHO topology (large α-cages). When this saturation uptake is expressed per mass of adsorbent, however, it yields a value of 31.0 mmol.g$^{-1}$, very close to that of ZIF-8 (29.1 mmol.g$^{-1}$). The heat of adsorption of the first molecules adsorbed is around 75 kJ/mol, much larger than the bulk vaporization enthalpy, but dropping sharply at higher water loading. All these results show that ZIF-65 (RHO) has amphiphilic character, with some parts of its internal surface hydrophilic, and some parts hydrophobic. The water adsorbs first near the hydrophilic patches of the internal surface as seen on Fig. 5 (right panel), showing the density of adsorbed water at 3.5 kPa on the adsorption branch). Then, at higher pressure, these "patches" of adsorbed water are joined by the full filling

of the pores, as demonstrated in earlier work on chemically heterogeneous (or "nanopatterned") pores.[51]

We then studied a different ZIF, ZIF-90, for which no experimental data is available. Figure 6 presents the water adsorption-desorption isotherms of this material. The adsorption isotherm in the gas phase of the ZIF-90 presents two successive regimes, like ZIF-65 (RHO). The heat of adsorption of the first molecules is around 65 kJ/mol, which is smaller than that of the ZIF-65 (RHO) but it is still much larger than the water vaporization enthalpy. Nevertheless, the heats of adsorption for the other molecules adsorbed in the ZIF-90 are around 55 kJ/mol which is higher than that of the ZIF-65 (RHO). The saturation uptake is around 80 molecules per unit cell (26.1 mmol.g$^{-1}$) and it is comparable to what we obtained for the ZIF-8 and its variants. We conclude that, due to the high affinity of the water molecules for the aldehyde group, ZIF-90 is strongly hydrophilic material. The two subsequent parts seen in the isotherms correspond to (i) strong adsorption sites featuring water–aldehyde hydrogen bonds; (ii) filling of the rest of the pore space. This is partly similar to ZIF-65 (RHO), but with the second part being less steep and without hysteresis.

In order to shed light into the nature of the strong adsorption sites of ZIF-90 at the molecular level, we have calculate the four water–aldehyde RDFs for water adsorbed at 0.2 kPa, i.e. in the low-pressure regime. The strong peak in the $O_{aldehyde}$–$H_{water}$ and $O_{aldehyde}$–$O_{water}$ RDFs, at 1.6 Å and 2.5 Å respectively, indicate the formation of ZIF-90–water hydrogen bonds where the H-bond donor is the carbonyl oxygen. There is, moreover, a strong rotational ordering of the water molecule: the difference between the two peaks, of 0.9 Å, corresponds almost exactly to the water O–H distance of 1.0 Å. This strong hydrogen bond, together with the presence of the neighboring H atom of the aldehyde group, creates a very favorable configuration for the water molecule, resulting in a very hydrophilic material.

Finally, in a study on $CO_2$ adsorption in ZIFs, Amrouche et al.[44] noted that the materials' affinity for $CO_2$ correlates well with the presence of polar functional groups on the linkers, which they quantified by the dipole moment of the ZIF linkers. Our results in this work confirm this correlation, water being a polar molecular fluid like $CO_2$ (water because of its dipole moment, carbon dioxide due to its large quadrupole moment).

### 3. Effect of pore geometry: ZIF-65 (SOD) and ZIF-7

Finally, we tried to gauge the effect of pore geometry on the water adsorption properties, in both hydrophilic and hydrophobic materials. First, we discuss the differences between ZIF-65 RHO and SOD. We present the water isotherms of the ZIF-65 (SOD) in the Figure 7. The isotherms and heat of adsorption indicate the hydrophilic nature of the material. Contrary to the ZIF-65 (RHO) which presents two gentles steps, the ZIF-65 (SOD) material presents a type I adsorption isotherm at low pressure, followed by a vertical step with hysteresis at higher pressure, though still in the gas phase. Around $P$ = 2 kPa, a plateau of 6 molecules per unit cell (2.2 mmol.g$^{-1}$) is observed, which corresponds to the adsorption of one molecule per 4-ring site, as depicted in Figure 8. This site is favorable, with two –NO$_2$ groups in close vicinity of the water molecule. The vertical transition at a pressure of 2.5 kPa corresponds to the filling of the rest of the pore volume, with a plateau at around 80 molec./u.c. (29.6 mmol.g$^{-1}$), as in other ZIFs of SOD topology and the ZIF-65 (RHO).

Finally, we studied the water adsorption properties of the ZIF-7 material, which presents the same topology and chemistry as ZIF-8: SOD topology, purely aromatic linkers and no hydrophilic functional groups. Figure 9 reports the water adsorption-desorption isotherms of the ZIF-7 material, and we observe a very different picture for ZIF-7 compared to ZIF-8.

Indeed, both the isotherms and heat of adsorption indicate a hydrophilic nature of ZIF-7, whereas ZIF-8 is hydrophobic and features no water adsorption in the gas phase. This difference in behavior is due to the structure of the material: while of the same SOD topology as ZIF-8, the geometry of ZIF-7 presents distortions from the ideal sodalite geometry, with two types of hexagonal windows (Figure 10). In a recent study, Aguado et al. also observed that the adsorption properties of the ZIF-7 and of the ZIF-8 are different for carbon dioxide isotherms at 303 K, highlighting the importance of small differences in geometry.

The amount of water adsorbed at saturation in the gas phase in ZIF-7 ($P = P_{sat}$) is 12 molecules per unit cell (2.2 mmol / g), and corresponds to the adsorption of two molecules in the smaller pockets (formed by the benzene rings of the linkers of a distorted 6-ring windows). These smaller pockets correspond to hydrophilic sites, due to their small size maximizing the interaction between adsorbed water molecule and neighboring aromatic rings by dispersive interactions (Figure 11). Up to 4 water molecules can be adsorbed in each of these hydrophilic sites, in a planar configuration depicted in Figure 11. The rest of the pore space of ZIF-7, i.e. the pockets centered on undistorted 6-rings, is highly hydrophobic: it only fills up with water upon liquid intrusion in the pressure range 200–800 MPa (not shown here). It also accommodate up to 4 water molecules, but they are in a tetrahedral arrangement (Figure 11) with weaker water–MOF interactions. We thus see how a seemingly small difference in pore size and geometry, at a given topology, can have phenomenal consequences on water adsorption properties.

In order to confirm that the changes in adsorption properties were due to geometry, and not to possible differences in polarity between ZIF-7's benzimidazolate linker and the linker of the parent MOF ZIF-8, we have calculated the dipole moments of the linkers. We find a value of 1.38 D for *bim* (ZIF-7), which is similar to that of *mim* (ZIF-8's linker; 1.25 D), and much lower than the dipole moment of the other hydrophilic ZIFs: 2.87 for *ica* (ZIF-90) and 3.12 for *nim* (ZIF-65).[52] The specific behavior of ZIF-7 is thus rooted in its particular geometry.

### 4. Generalization and comparison with zeolites and other MOF families

We now turn our attention to the systems studied as a *family* of related materials, and try to give a general understanding of how geometry and functionalization affect water adsorption properties. Starting from a parent hydrophobic material, we then look at the effect of functionalization or changes in pore geometry. We have shown, in the past, that this approach enables one to rationalize the evolution of adsorption behavior in families of materials such as hydrophobic MOF Al(OH)(naphthalenedicarboxylate),[53] $CO_2$ adsorption in IRMOFs,[54] and water adsorption various zeolites.[55,56]

In the family of ZIFs related to the widely studied ZIF-8, the *parent* compound of the family is SALEM-2, the unfunctionalized Zn(imidazolate)$_2$ structure of SOD topology. SALEM-2 is hydrophobic, with water intrusion happening in liquid phase and following a type V isotherm, with vertical transitions and a hysteresis both characteristic of a first-order vapor-to-liquid transition. This is similar to that found in hydrophobic (pure silica) zeolites with pores of similar size,[55] i.e. relatively large pores by zeolites' standards. By functionalizing the material with hydrophobic (or neutral) groups like –CH$_3$ and –Cl, we change only slightly the water–MOF interactions and pore volume, hence only changing slightly the intrusion and extrusion pressures, but retaining the fully hydrophobic behavior. If we introduce a polar group such as –NO$_2$, as in ZIF-65 (SOD), we increase the water–MOF interactions in the vicinity of the NO$_2$ ($\Delta H_{ads}$ = 52.5 kJ/mol), but keep an overall hydrophobic MOF. At low pressure, we thus observe a small uptake of water in specific hydrophilic sites. This reduces the hydrophobicity of the material, to the point that adsorption now takes place in the gas phase (at $P < P^0$). However, the adsorption-desorption transition remains sharp and hysteretic. This behavior has been observed before in hydrophobic zeolites with heterogeneous nanopores (presence of silanol nests). It was characterized by Cailliez et al. as a *weak defect* of the hydrophobic structure,[57] i.e. a hydrophilic defect which creates a new adsorption site, and shifts the adsorption transition while retaining the overall hydrophobic character of the material.

In contrast with that situation, we also found in our study of ZIFs evidence of behavior defined by Cailliez as *strong defects*. When a stronger hydrophilic group is added to the material, such as –COH in ZIF-90, the water–MOF interactions become so strong compared to water–water interactions ($\Delta H_{ads}$ = 80 kJ/mol, compared to $\Delta H_{vap}$ = 44 kJ/mol) that the nature of the adsorption transition itself is changed: the material is fully hydrophilic, with a reversible type I adsorption-desorption isotherm. Thus, the conclusions drawn in these earlier work in the case of inorganic materials and other MOF systems apply equally to the ZIF family of materials, and allow one to rationalize the impact of internal pore surface functionalization on water adsorption properties.

### 5. Nature of the adsorbed phase at saturation

Finally, we were interested in the nature of adsorbed phase at saturation in each of the materials studied. In each case, we have calculated the radial distribution functions for the adsorbed water at saturation (we used a pressure of 250 MPa for comparison purposes). We show all the O–O RDFs in Figure 12. We observe that the structure of adsorbed water at saturation is essentially identical for all ZIFs except ZIF-7, whose adsorbed water is more structured to fragmentation of the pore space in two disjoint spaces. Water adsorbed in all other MOFs in the liquid phase has a structure quite similar to bulk water. The same is true based on the O–H and H–H RDFs (not shown here). We thus conclude that water adsorbed in ZIFs at saturation retains bulk liquid-like characteristics, owing to the quite large pore space of these materials compared to zeolites, where the local water density is significantly lower than bulk, even at saturation (e.g., density of 0.6 g.cm$^{-3}$ in silicalite-1[58]).

## IV. Conclusions

By using molecular simulation on different members of the ZIF family, we demonstrated that topology, geometry, and linker functionalization can drastically affect the water adsorption properties of these materials, tweaking the water–ZIF interactions from hydrophobic to hydrophilic, and even featuring dual amphiphilic materials whose pore space features both hydrophobic and hydrophilic regions. We show that adequate functionalization of the linkers allow one to tune the host–guest interactions, which could be used to design new materials with specific water adsorption properties. In particular, the change in pore size in hydrophobic materials could be used to tune the water intrusion and extrusion pressures, which are key parameters of the energetic performance for energy storage and shock-absorber behavior. These results are in keeping with earlier work on water adsorption in a family of hydrophobic MOFs: Al(OH)(1,4-naphthalenedicarboxylate) and derivatives.

Moreover, we showed that depending on the type functionalization introduced in an initially hydrophobic material (ZIF-8), various degrees of hydrophilicity could be obtained, with a gradual evolution from a type V adsorption isotherm in the liquid phase, to a type I isotherm in the gas phase. A similar behavior was observed in families of hydrophobic all-silica zeolites, with hydrophilic "defects" of various strength: silanol nests, which are weak defects leading to local adsorption but no overall change in the type of the isotherm; extraframework cations, which are strongly hydrophilic and lead to a global change in the nature of the adsorption. As this work shows, the lessons learnt in water adsorption in hydrophobic zeolites and their derivatives, including hydrophilic materials, can serve as a guide to better understanding the water adsorption properties of the families of new hydrophobic hybrid organic–inorganic materials, such as ZIFs, and the influence of small changes in geometry, topology, and functionalization on the properties of these new materials.

## Acknowledgments

Funding from the Agence Nationale de la Recherche under project "SOFT-CRYSTAB" (ANR-2010-BLAN-0822) is acknowledged.

## Notes and references

[a] CNRS and Chimie ParisTech, 11 rue Pierre et Marie Curie, 75005 Paris, France. E-mail: fx.coudert@chimie-paristech.fr

[b] Département de Chimie, École Normale Supérieure, CNRS-ENS-UPMC, 24 rue Lhomond, 75005 Paris, France.

† Electronic Supplementary Information (ESI) available: force field parameters, water–ZIF-90 radial distribution functions, structures of ZIF-Cl and SALEM-2. See DOI: 10.1039/b000000x/

| Framework | Linker | Topology | Space group | Unit cell parameters | Unit cell volume | Ref. |
|---|---|---|---|---|---|---|
| **ZIF-8** | 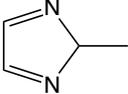 | SOD | $I\,\bar{4}\,3\,m$ | $a = 16.99$ Å | 4907.1 Å$^3$ | 7 |
| **ZIF-7** | 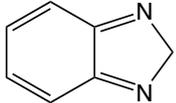 | SOD | $R\,\bar{3}$ | $a = 22.22$ Å<br>$c = 16.08$ Å | 6878.2 Å$^3$ | 36 |
| **ZIF-90** | 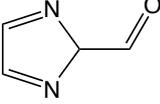 | SOD | $I\,\bar{4}\,3\,m$ | $a = 17.27$ Å | 5152.2 Å$^3$ | 32 |
| **ZIF-65 (SOD)** | 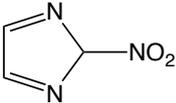 | SOD | $I\,\bar{4}\,3\,m$ | $a = 17.27$ Å | 5152.2 Å$^3$ | 28 |
| **ZIF-65 (RHO)** | 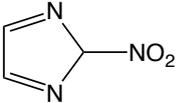 | RHO | $I\,\bar{4}\,3\,m$ | $a = 29.03$ Å | 24465.3 Å$^3$ | 28 |
| **ZIF-Cl** | 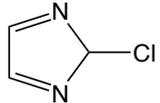 | SOD | $I\,\bar{4}\,3\,m$ | $a = 17.13$ Å | 5020.7 Å$^3$ | 35 |
| **SALEM-2** | 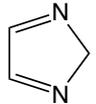 | SOD | $I\,\bar{4}\,3\,m$ | $a = 17.01$ Å | 4925.2 Å$^3$ | 33 |

**Table 1.** Main characteristics of the ZIFs studied, including ligand structure, name, crystalline system, space group, lattice parameters, and lattice volume.

| Framework | ZIF-8 | ZIF-7 | ZIF-90 | ZIF-65 (SOD) | ZIF-65 (RHO) | ZIF-Cl | SALEM-2 |
|---|---|---|---|---|---|---|---|
| **H$_{ads}$ (kJ/mol)** | 15.5 | 39.5 | 79.7 | 52.5 | 69.5 | 17.8 | 13.6 |

**Table 2.** Adsorption enthalpy for the first water molecule (limit of zero loading) in each of the materials studied.

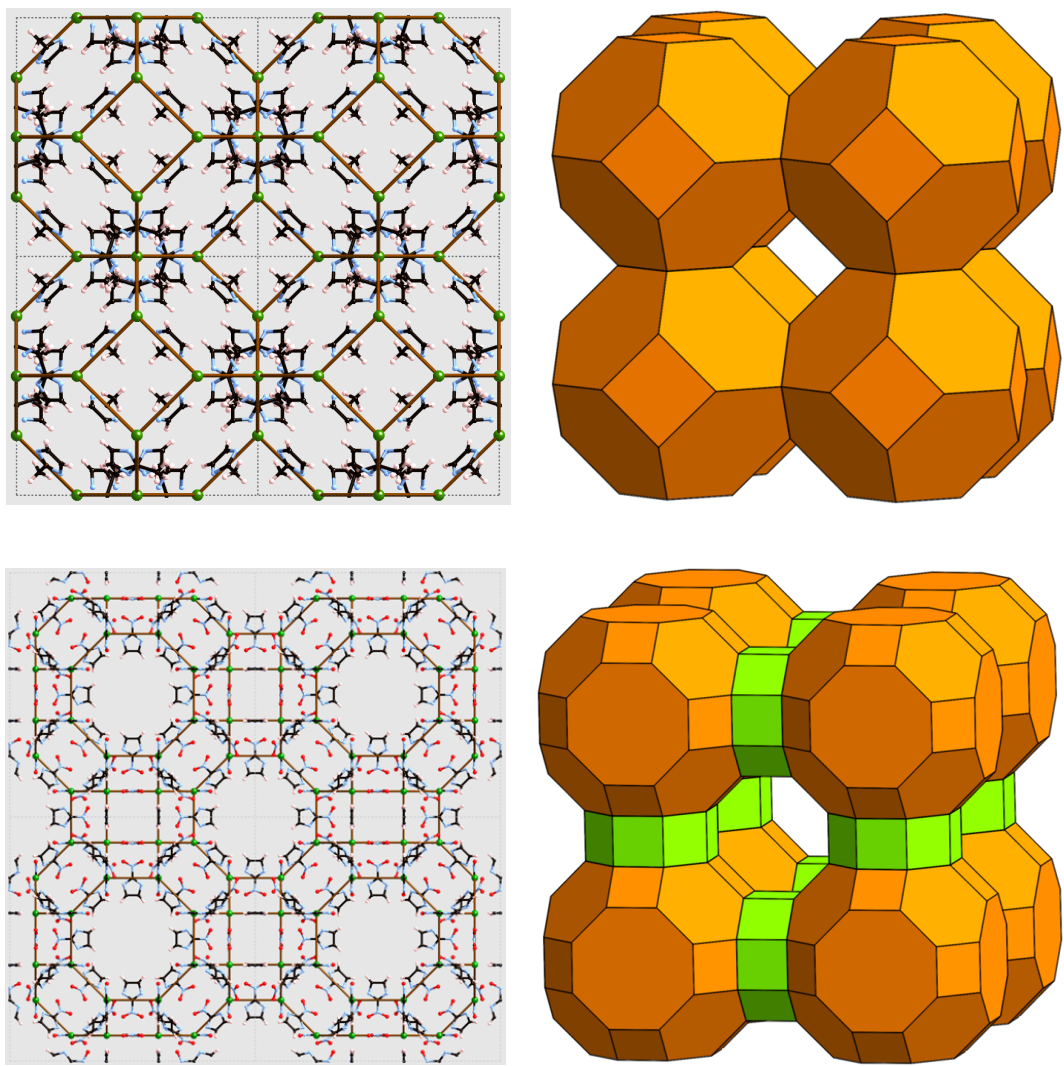

**Figure 1.** Top: representation of a 2 × 2 × 2 supercell of the ZIF-8 structure, highlighting the SOD topology (schematized on the right) of the framework with brown lines drawn between neighboring $Zn^{2+}$ ions (in green). Bottom left: representation of the unit cell of ZIF-65 (RHO), highlighting the RHO topology (schematized on the right) of the framework with brown lines drawn between neighboring $Zn^{2+}$ ions (in green).

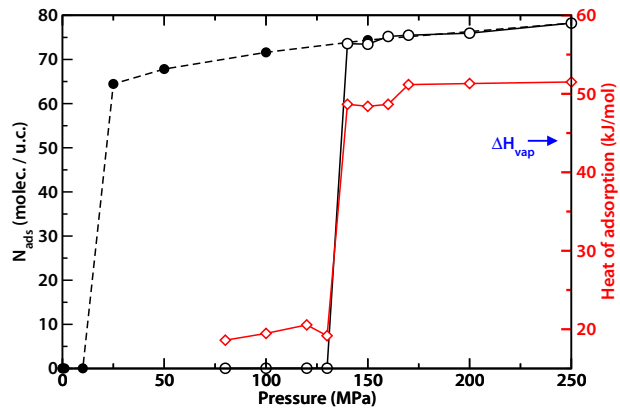

**Figure 2.** Simulated adsorption (black open symbols) and desorption (black full symbols) isotherms of water intrusion in ZIF-8 at 300 K. The red curve represents the heat of adsorption (scale on the right) and the blue arrow indicates the bulk water enthalpy of vaporization at 300 K.

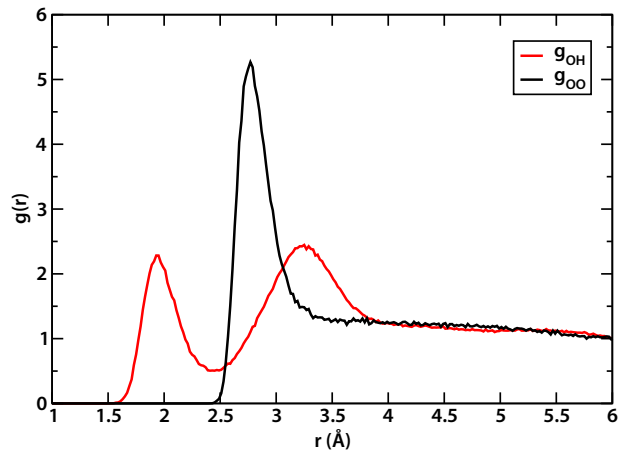

**Figure 3.** The O–O radial distribution function (black curve) and the O–H radial distribution function (red curve) for water molecules in the ZIF-8 at 250 MPa.

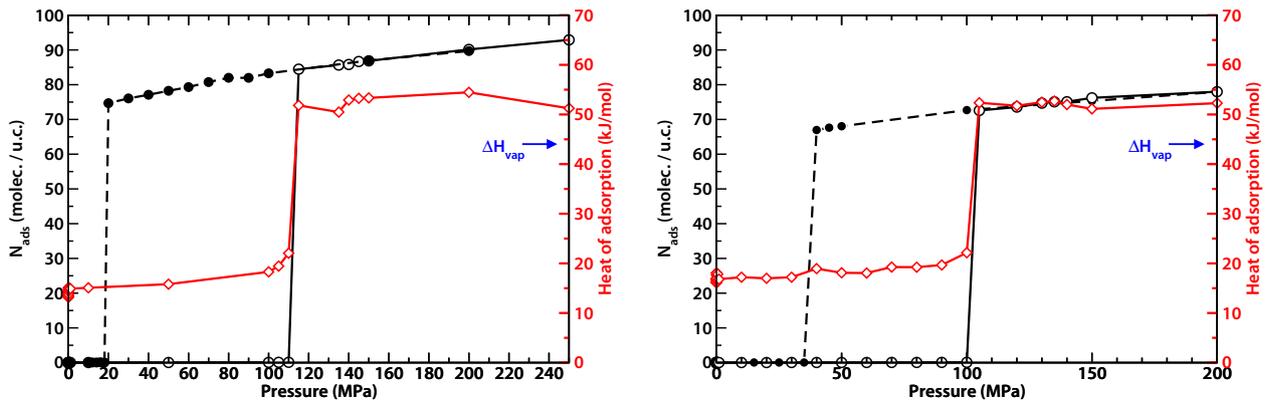

**Figure 4.** Simulated adsorption-desorption isotherms of water at 300 K in SALEM-2 (left) and in ZIF-Cl (right). Open symbols for adsorption and full symbol for desorption. Red curve represents the heat of adsorption.

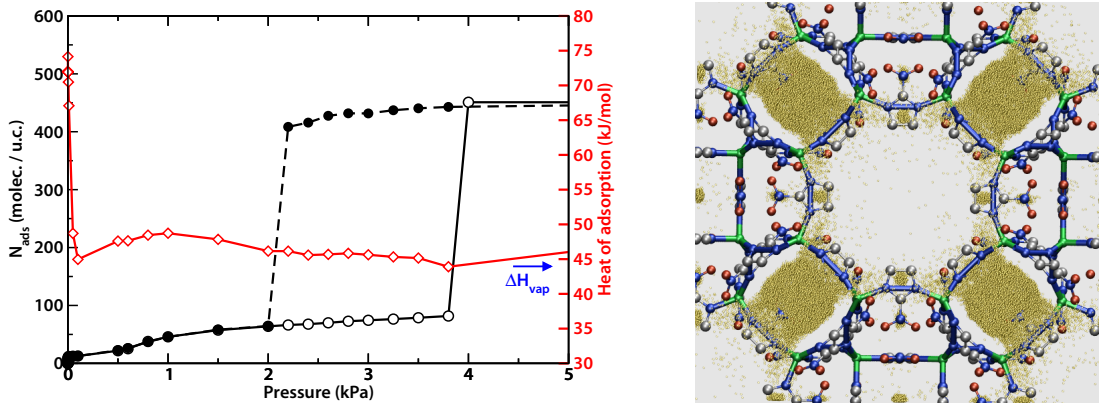

**Figure 5.** Left: Simulated adsorption-desorption isotherms of water at 300 K in the ZIF-65 (RHO) material. Open symbols for adsorption and full symbol for desorption. Red curve represents the heat of adsorption. Right: density of water (yellow) in the pores of ZIF-65 (RHO) at 3.5 kPa on the adsorption branch.

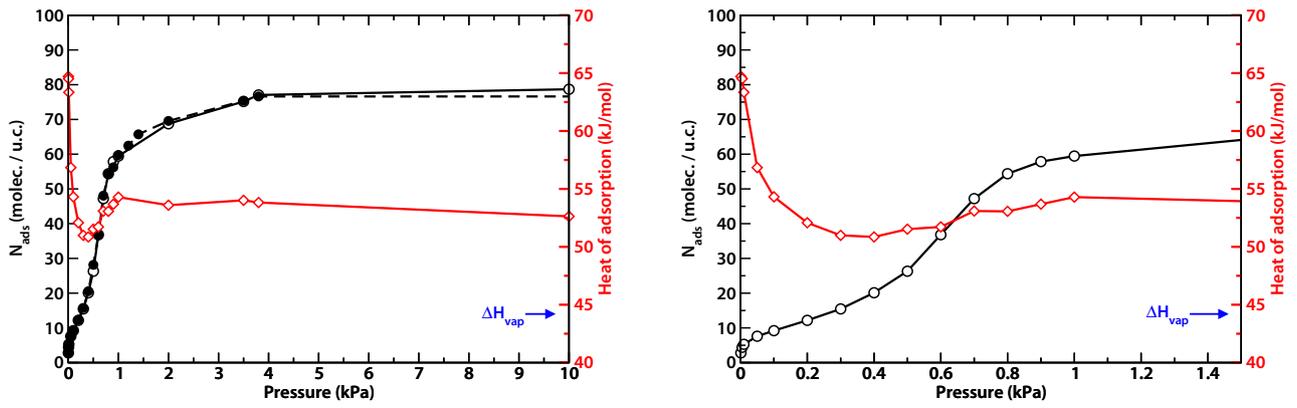

**Figure 6.** Simulated adsorption-desorption isotherms of water at 300 K in the ZIF-90 material (on the left), zoom in the pressure ranging from 0 to 1.6 kPa (on the right). Open symbols for adsorption and full symbol for desorption. Red curve represents the heat of adsorption.

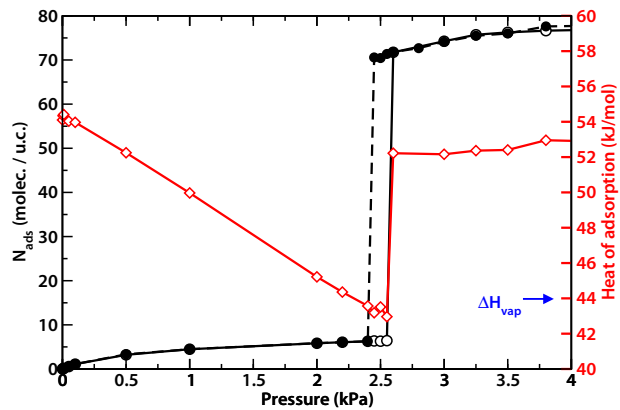

**Figure 7.** Simulated adsorption-desorption isotherms of water at 300 K in the ZIF-65 (SOD) material. Open symbols for adsorption and full symbol for desorption. Red curve represents the heat of adsorption.

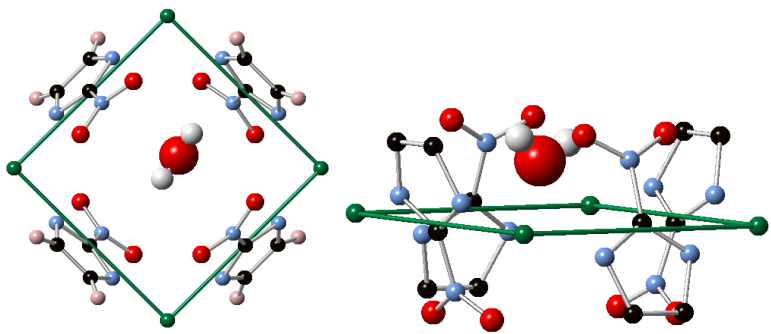

**Figure 8.** Snapshot of the 4-ring water adsorption site of the ZIF-65 (SOD) material in the gas phase.

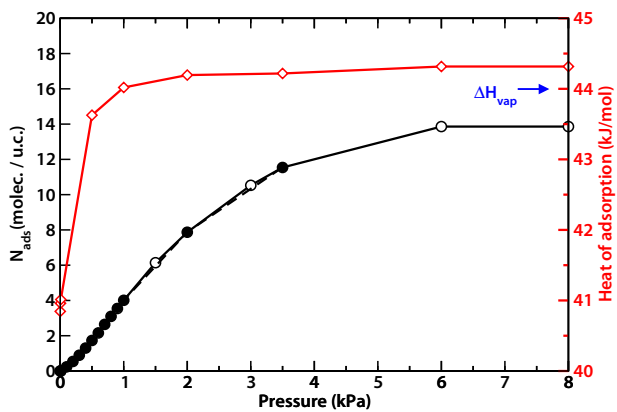

**Figure 9.** Simulated adsorption-desorption isotherms of water at 300 K in the ZIF-7 material, at low pressure. Open symbols for adsorption and full symbol for desorption. Red curve represents the heat of adsorption.

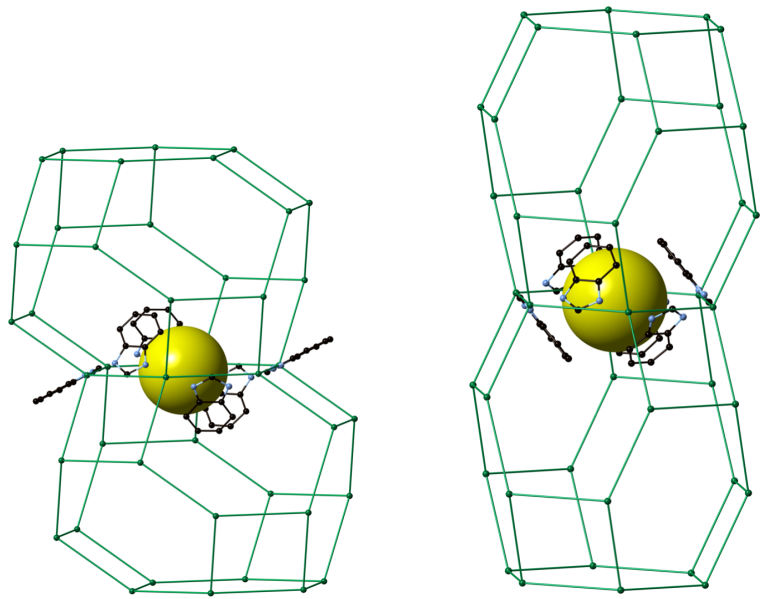

**Figure 10**. Snapshots of the two different pores of the ZIF-7 material: smaller on the left, larger on the right. The yellow spheres represent the porous volume of the pores. The two joined sodalite cages of the SOD framework are represented in green wire.

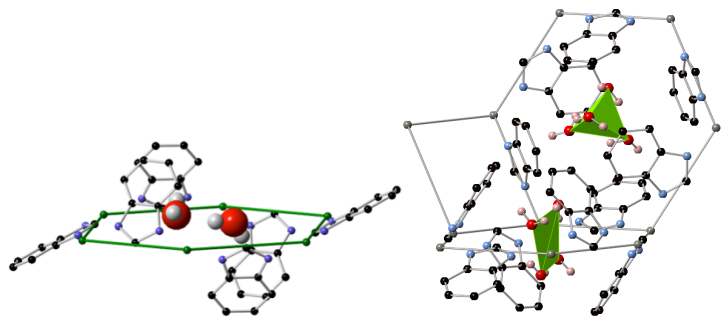

**Figure 11.** Pictures of the water adsorption sites of the ZIF-7 material. On the left for low water pressure ($P = 1$ kPa) and on the right for high water pressure ($P = 800$ MPa).

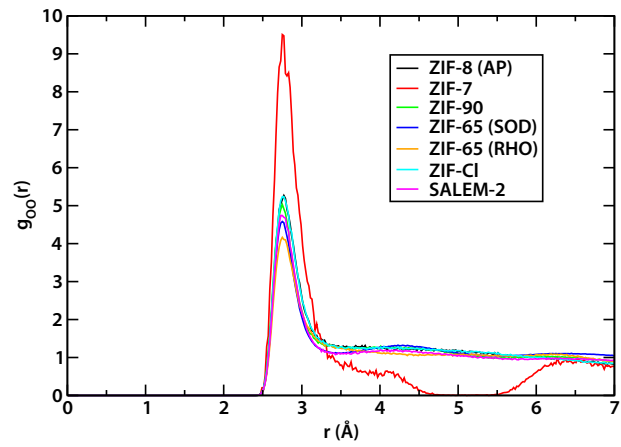

**Figure 12.** Radial distribution functions of oxygen atoms of water molecules adsorbed at 250 MPa in ZIF-8, ZIF-7, ZIF-90, ZIF-65(SOD), ZIF-65(RHO), ZIF-Cl, and SALEM-2.